\documentclass[floatfix,aps,amsmath,nofootinbib,twocolumn,10pt]{revtex4}

\usepackage{listings}
\usepackage{graphicx}
\usepackage{bm}
\usepackage{rotating}
\usepackage{array}
\usepackage{amsmath}
\usepackage{amssymb} 
\usepackage{mathrsfs} 
\usepackage{cancel}
\usepackage{subfig}
\usepackage{float}
\usepackage{caption}

\def\({\left(}
\def\){\right)}
\def\[{\left[}
\def\]{\right]}

\def\e{\begin{equation}}
\def\q{\end{equation}}
\def\m{\begin{eqnarray}}
\def\n{\end{eqnarray}}

\begin{document}
\title{Frequency Modulation of Gravitational Waves by Ultralight Scalar Dark Matter}
\author{Ke Wang$^{1,2,3}$}
\thanks{Corresponding author: {wangkey@lzu.edu.cn}}
\author{Yin Zhong$^{2,3}$}
\affiliation{$^1$Institute of Theoretical Physics $\&$ Research Center of Gravitation, Lanzhou University, Lanzhou 730000, China}
\affiliation{$^2$Key Laboratory of Quantum Theory and Applications of MoE, Lanzhou University, Lanzhou 730000, China}
\affiliation{$^3$Lanzhou Center for Theoretical Physics $\&$ Key Laboratory of Theoretical Physics of Gansu Province, Lanzhou University, Lanzhou 730000, China}

\date{\today}

\begin{abstract}
The oscillating pressure of the ultralight scalar dark matter (DM) can induce the oscillation of the local gravitational potential. Similar to the time-dependent frequency shift for the pulse signals of pulsars, the oscillation of the local gravitational potential can induce a time-dependent frequency shift (or frequency modulation) for quasi-monochromatic gravitational wave (GW) signals from galactic white dwarf (WD) binaries. To make this effects detectable, we suppose that some galactic WD binaries are located in the DM clumps/subhalos where the energy density of DM is about eight orders of magnitude higher than that at the position of the Earth. Turning to the fisher information matrix, we find that the amplified GW frequency modulation induced by the ultralight scalar DM with mass $m=1.67\times10^{-23}-4.31\times10^{-23}[{\rm eV}/c^2]$ can be detected by LISA.
\end{abstract} 

\maketitle


\section{Introduction}
\label{sec:intro}
The rotational properties of galaxies~\cite{Rubin:1982kyu}, the evolution of large scale structure~\cite{Davis:1985rj} and the gravitational lensing observations~\cite{Clowe:2006eq} are considered to be the direct empirical proofs of the existence of dark matter (DM). Based on the standard Lambda cold DM ($\Lambda$CDM) cosmological model, the latest cosmic microwave background (CMB) observations~\cite{Planck:2018vyg} further suggests about $26\%$ of the energy density in the Universe comes from CDM today. However, as one of the most promising candidates for CDM, the weak interacting massive particles (WIMPs) grounded on supersymmetric theories of particle physics still have not been detected~\cite{PandaX-II:2016vec,LUX:2015abn,ATLASCMS,AMS:2014bun,Fermi-LAT:2011baq}. Moreover, primordial black holes which can also serve as CDM~\cite{Carr:2016drx} still have not been identified. These null results accompanied by CDM's failure on sub-galactic scales~\cite{Primack:2009jr} imply that the standard CDM model may be not the final answer.

The de Broglie wavelength of the ultralight non-interacting particles with mass $\sim10^{-22}[{\rm eV}/c^2]$ is comparable to astrophysical scales $\sim60[{\rm pc}]$. As a result, ultralight particles can smooth out the inhomogeneities on small scales and prevent sub-galactic structures forming. According to this effect, an alternative candidate for DM is proposed. This ultralight DM (ULDM) can not only behave as CDM on large scales but also avoid the CDM small scale crises~\cite{Hu:2000ke}. Also due to the wave nature, the pressure of ULDM is coherently oscillating. And the oscillation of the pressure can induce the oscillation of the metric in the DM halo. Similar to gravitational waves (GWs), the time-dependent perturbations of the background metric induced by ULDM can also change the pulse arrival time of the pulsar and be detected by pulsar timing arrays (PTAs). The simplest cases are that ULDM is the ultralight axion-like scalar particles~\cite{Khmelnitsky:2013lxt,Marsh:2015xka,Kato:2019bqz}. After that, the pulsar timing residual induced by ultralight vector particles was investigated~\cite{Nomura:2019cvc}. Recently, the pulsar timing residual induced by ultralight tensor particles is also investigated~\cite{Wu:2023dnp}.

Besides detecting ULDM by PTAs, many other detection methods of ULDM are proposed. Similar to the GW detection, for example, the direct detection of ULDM wind by space-based laser interferometers such as Laser Interferometer Space Antenna (LISA)~\cite{LISA:2017pwj} has been estimated~\cite{Aoki:2016kwl}. ULDM can also affect orbital motions of astrophysical objects in the galaxy and be detected indirectly~\cite{Blas:2016ddr,Boskovic:2018rub}.
Moreover, the black hole superradiant instability from ULDM can also constrain its mass~\cite{Brito:2020lup}.

In this paper, we propose a new novel detection method of ULDM. Similar to the time-dependent frequency shift for the pulse signals of pulsars, the oscillation of the local gravitational potential can induce a time-dependent frequency shift (or frequency modulation) for quasi-monochromatic GW signals from galactic white dwarf (WD) binaries. Although there are about $10^7$ WD binaries in the Milky Way~\cite{Nelemans:2001hp}, the number of WD binaries with chirp mass measured by LISA is only about $1000$~\cite{Lamberts:2019nyk}. Here we assume some WD binaries with chirp mass measured by LISA are located in the DM clumps/subhalos~\footnote{On the one hand, the de Broglie wavelength of free ULDM with mass $\sim10^{-22}[{\rm eV}/c^2]$ is much larger than the size of DM clumps/subhalos $r\sim2[{\rm pc}]$. 
On the other hand, the observations of GD-1 stellar stream do favor such DM clumps/subhalos.
We solve this tension by assuming that there is an additional local potential well $V(r)\approx\frac{1}{2}m\omega_0^2r^2+...$ at the location of DM clump/subhalo. Then the size of the DM clump/subhalo will be $r=\sqrt{\frac{\hbar}{m\omega_0}}$. And $r\sim2[{\rm pc}]$ just needs a very flat potential well with $\omega_0\sim10^{-10}{\rm[Hz]}$ where the of behavior of ULDM will be very similar to that of ones outside the DM clump/subhalo.
Since such potential wells are formed coincidentally and the number of them is small ($\sim100$~\cite{Mirabal:2021ayb}) in the Milky Way, the existence of them will not affect the statistical fact that ULDM suppresses the mass power spectrum on small scales~\cite{Hui:2021tkt}}. 
The DM clumps/subhalos with mass $m\sim10^7M_{\odot}$ and size $r^3\sim10[{\rm pc}^3]$ can serve as massive perturbers to explains many of the observed stream features in GD-1 stellar stream, such as the spurs and the gaps~\cite{Bonaca:2018fek,Mirabal:2021ayb}. As a result, the GW frequency modulation induced by the ultralight scalar DM will be amplified by about eight orders of magnitude compared to the same effect taking place at the position of the Earth. Taking this mechanism into consideration and turn to the fisher information
matrix, we can estimate the detection of the ultralight scalar DM by LISA~\cite{LISA:2017pwj}. 

This paper is organized as follows.
In section~\ref{sec:MID}, we estimate the detection of GW frequency modulation by LISA model-independently.
In section~\ref{sec:MD}, we forecast the constraints on ultralight scalar DM by the detection of GW frequency modulation.
Finally, a brief summary and discussions are included in section~\ref{sec:SD}.

\section{Detection of GW Frequency Modulation by LISA}
\label{sec:MID}
\subsection{GW Signals and Detector}
\label{ss:S&D}
Galactic WD binaries are supposed to be quasi-monochromatic GW sources. Therefore, the GW signal from them in their own frame is defined as:
\begin{align}
h_{+}(t)=&(\mathcal{A}+\delta\mathcal{A})(1+\cos^2\iota)\cos(\phi(t)),\\
h_{\times}(t)=&-2(\mathcal{A}+\delta\mathcal{A})\cos\iota\sin(\phi(t)),
\end{align}
where the involved derived parameters including the dimensionless amplitude $\mathcal{A}$, the phase $\phi$, the chirping frequency $\Dot{f_0}$ and the chirp mass $\mathcal{M}$ are defined as:
\begin{align}
\label{eq:am}
\mathcal{A}=&\frac{2(G\mathcal{M})^{5/3}(\pi f_0)^{2/3}}{c^4d},\\
\phi(t)=&2\pi f_0t+\pi\Dot{f_0}(1+\delta\Dot{f})t^2+\phi_0,\\
\Dot{f_0}=&\frac{96}{5}\pi^{8/3}\left(\frac{G\mathcal{M}}{c^3}\right)^{5/3}f_0^{11/3},\\
\mathcal{M}=&\frac{(m_1m_2)^{3/5}}{(m_1+m_2)^{1/5}}.
\end{align}
Above derived parameters are further dependent on the primary and secondary WD masses $m_1$ and $m_2$, the luminosity distance to the binary $d$, the frequency of GW $f_0$, the orbital inclination $\iota$ and the initial GW phase $\phi_0$. Besides these common parameters, we introduce two deviation parameters $\delta\mathcal{A}$ and $\delta\Dot{f}$ to characterize the amplitude modulation and the frequency modulation during GW propagation. In the following discussion, we will consider a specific galactic WD binary, whose parameters are listed in Tab.~\ref{tb:para}.
\begin{table}
\captionsetup{justification=raggedright}
\caption{Parameters of one specific WD binary and LISA constellation. The parameters in the first row are necessary to obtain the GW signal in the source frame. The parameters in the second row are necessary to obtain the GW response of the TDI observables. The parameters in the third row are newly introduced or derived parameters.}
\label{tb:para}
\begin{tabular}{|cccccc|}
\hline
$m_1[M_{\odot}]$ & $m_2[M_{\odot}]$ & $d[{{\rm kpc}}]$ & $f_0[{\rm Hz}]$ & $\iota[{\rm rad}]$ & $\phi_0[{\rm rad}]$\\
\hline 
$1$ & $1$ & $1$ & $1\times10^{-3}$ & $\frac{\pi}{4}$ & $0$\\
\hline
\hline
$\psi[{\rm rad}]$ & $\beta[{\rm rad}]$ & $\lambda[{\rm rad}]$ & $L_i[{\rm km}]$ & $T[{\rm year}]$ & $e$\\
\hline
$\frac{\pi}{4}$ & $-\frac{\pi}{4}$ & $\frac{\pi}{4}$ & $2.5\times10^{6}$ & $4$ & $0.00964838$\\
\hline 
\hline
$\delta\mathcal{A}$ & $\delta\Dot{f}$ & $\mathcal{A}$ & $\Dot{f_0}[{\rm Hz}^{2}]$ & $\mathcal{M}[M_{\odot}]$ & SNR\\
\hline 
$0$ & $0$ & $4.7\times10^{-22}$ & $4.6\times10^{-18}$ & $0.87$ & $122$\\
\hline
\end{tabular}
\end{table}

\begin{figure*}[]
\begin{center}
\subfloat{\includegraphics[width=0.33\textwidth]{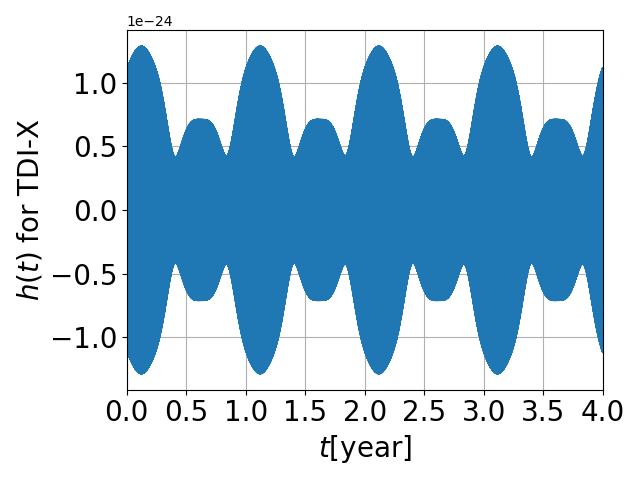}}
\subfloat{\includegraphics[width=0.33\textwidth]{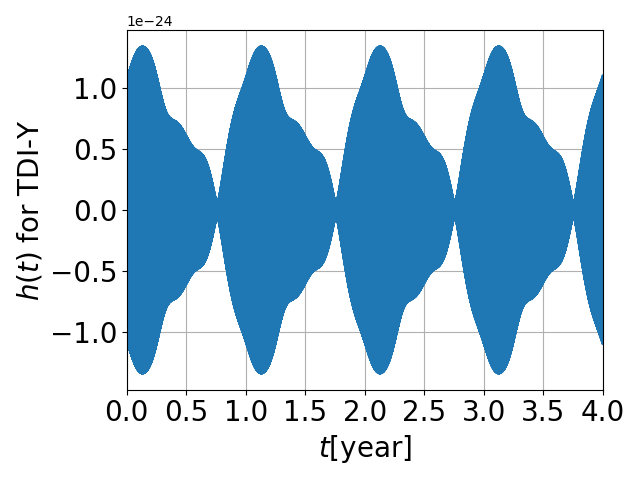}}
\subfloat{\includegraphics[width=0.33\textwidth]{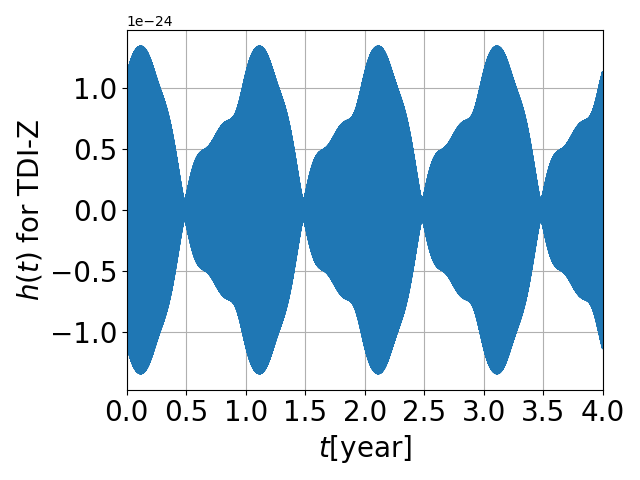}}
\end{center}
\captionsetup{justification=raggedright}
\caption{The GW response of the second-generation TDI observables: X, Y, Z. The GW signal in the source frame is given by the parameters in the first row of Tab.~\ref{tb:para} and the detector’s (LISA) response is determined by the parameters in the second row of Tab.~\ref{tb:para}  }
\label{fig:h}
\end{figure*}

\begin{figure*}[]
\begin{center}
\subfloat{\includegraphics[width=0.33\textwidth]{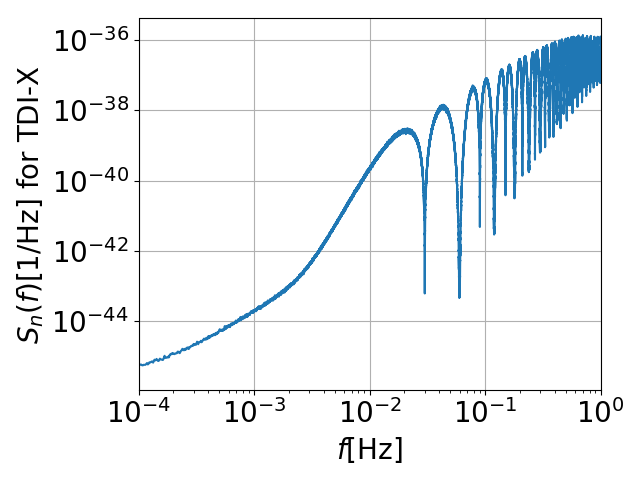}}
\subfloat{\includegraphics[width=0.33\textwidth]{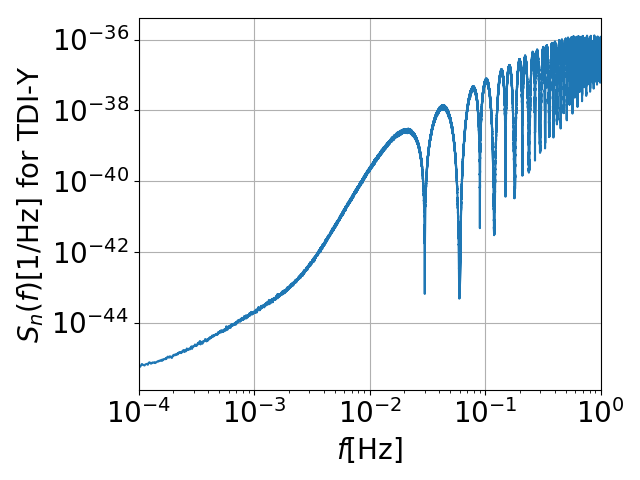}}
\subfloat{\includegraphics[width=0.33\textwidth]{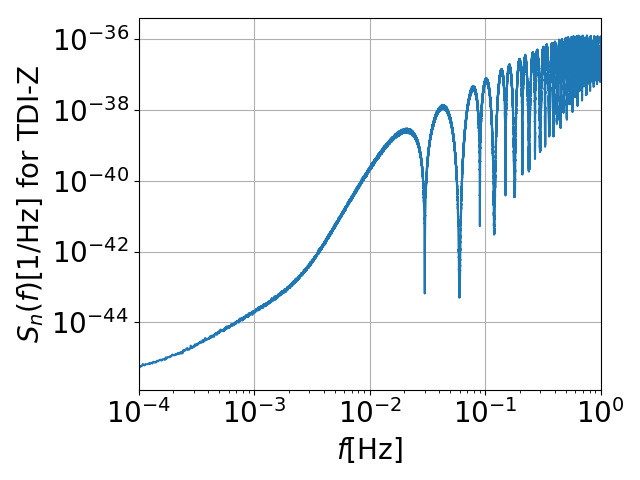}}
\end{center}
\captionsetup{justification=raggedright}
\caption{The noise PSD derived from the response of the second-generation TDI observables: X, Y, Z to the combination of fundamental LISA noises: laser frequency noise, proof-mass noise and optical-path noise, where $S_n(f_0)=2.12\times10^{-44}[{\rm Hz^{-1}}]$, $S_n(f_0)=2.07\times10^{-44}[{\rm Hz^{-1}}]$ and $S_n(f_0)=2.01\times10^{-44}[{\rm Hz^{-1}}]$ for X, Y, Z channel respectively.}
\label{fig:s}
\end{figure*}

Given the three armlengths of the LISA constellation~\cite{LISA:2017pwj} are $L_1=L_2=L_3=2.5\times10^{6}[{\rm km}]$, the mission lifetime of the LISA is $T=4[{\rm year}]$ and the eccentricity of the LISA spacecraft orbits in the Solar-system barycentric ecliptic coordinate system is $e=0.00964838$, we can calculate the GW response $h(t)$ of the second-generation Time Delay Interferometry (TDI) observables (e.g. X, Y, Z) for a GW source with the polarization angle $\psi$ at ecliptic latitude $\beta$ and ecliptic longitude $\lambda$. Meanwhile, we can calculate the response of the second-generation TDI observables (e.g. X, Y, Z) to the combination of fundamental LISA noises including laser frequency noise, proof-mass noise and optical-path noise $n(t)$ or its power spectral density (PSD) $S_n(f)$.
In this paper, we use the $\mathsf{Synthetic~LISA}$~\cite{Vallisneri:2004bn} (C++/Python2.x) to simulate the response of the second-generation TDI observables (e.g. X, Y, Z) to GW signals and noises, as shown in Fig.~\ref{fig:h} and Fig.~\ref{fig:s} respectively. One can also use other simulators such as $\mathsf{LISACode}$~\cite{Petiteau:2008zz} or the analytical formulations~\cite{Cutler:1997ta,Estabrook:2000ef,Cornish:2002rt,Prince:2002hp,Robson:2018ifk,Babak:2021mhe} to obtain the GW and noise responses.
Finally the output of LISA is
\begin{equation}
s(t)=h(t)+n(t).
\end{equation}
To implement the parameter estimation, we should tell $h(t)$ from $n(t)$. We can do that only for those sources whose signal-to-noise ratio (SNR) are high enough 
\begin{equation}
{\rm SNR}^2=\sum_{X,Y,Z}4{\rm Re}\left(\int_0^{\infty}df\frac{\tilde{h}^*(f)\tilde{h}(f)}{S_n(f)}\right).
\end{equation}
For quasi-monochromatic GW sources with the initial frequency $f_0$, we can use the Parseval’s theorem to re-write ${\rm SNR}^2$ in the time domain
\begin{equation}
{\rm SNR}^2=\sum_{X,Y,Z}\frac{2}{S_n(f_0)}\int_0^{T}dt~h(t)~h(t).
\end{equation}
In Tab.~\ref{tb:para}, we also list the SNR for our example of galactic WD binary.

\subsection{Fisher Information Matrix}
\label{ss:fish}
To forecast the constraints on parameters, in this paper, we will turn to the fisher information matrix
\begin{equation}
\label{eq:F}
\mathcal{F}_{ij}=\sum_{X,Y,Z}\frac{2}{S_n(f_0)}\int_0^{T}dt~\frac{\partial h(t;\bm{\theta})}{\partial \bm{\theta}_i}\frac{\partial h(t;\bm{\theta})}{\partial \bm{\theta}_j},
\end{equation}
where $\bm{\theta}$ is a vector consisting of $7$ free parameters
\begin{equation}
\nonumber
\bm{\theta}=\{\delta\mathcal{A},\delta\Dot{f},\iota,\phi_0,\psi,\beta,\lambda\}.
\end{equation}
The root mean square errors of these parameters are given
by
\begin{equation}
\sigma_i=\sqrt{(\mathcal{F}^{-1})_{ii}}.
\end{equation}
To numerically plug the GW responses of TDI observables $h(t;\bm{\theta})$ and the noise PSD $S_n(f)$ into Eq.~(\ref{eq:F}), we re-write the fisher information matrix calculation package for GW  detector networks $\mathsf{GWFISH}$~\cite{Dupletsa:2022scg} (Python3.x) in Python2.x and make it compatible with $\mathsf{Synthetic~LISA}$~\cite{Vallisneri:2004bn} (C++/Python2.x). Finally we use $\mathsf{GWFISH}$ to obtain the measurement uncertainties, as listed in Tab.~\ref{tb:err}.
\begin{table}
\captionsetup{justification=raggedright}
\caption{Measurement uncertainties obtained with $\mathsf{Synthetic~LISA}$ plugged into $\mathsf{GWFISH}$. The errors are at $68\%$ confidence level.} 
\label{tb:err}
\begin{tabular}{|cccc|}
\hline
$\delta\mathcal{A}$ & $\delta\Dot{f}$ & $\iota[{\rm rad}]$ & $\phi_0[{\rm rad}]$\\
\hline 
$0\pm0.041$ & $0\pm0.098$ & $\frac{\pi}{4}\pm0.049$ & $\frac{\pi}{4}\pm0.14$\\
\hline
\hline
$\psi[{\rm rad}]$ & $\beta[{\rm rad}]$ & $\lambda[{\rm rad}]$ & -\\
\hline
$\frac{\pi}{4}\pm0.070$ & $-\frac{\pi}{4}\pm0.0053$ & $\frac{\pi}{4}\pm0.0057$ & -\\
\hline
\end{tabular}
\end{table}
In this paper, we just care about $\sigma_{\delta\dot{f}}$ which is the error of $\delta\Dot{f}$. From the definition of $\delta\Dot{f}$, we know that the frequency modulation during GW propagation larger than $\sigma_{\delta\dot{f}}\Dot{f}=4.5\times10^{-19}[{\rm Hz}^{2}]$ will be detected by LISA.

\section{Forecasting the Constraints on Ultralight Scalar DM}
\label{sec:MD}
If we confirm GW signals are quasi-monochromatic in their source frame, they can be considered to be an unique probe during GW propagation. For example, the amplitude modulation taking place during GW propagation imply that there may be an evolving gravitational lens~\cite{Qiu:2022dya}. In this paper, we will investigate the frequency modulation induced by ultralight scalar DM during GW propagation.

Since the pressure of ultralight scalar DM is coherently oscillating in the galactic, the surrounding metric has the following form
\begin{equation}
ds^2=(1+2\Phi(\mathbf{x},t))^2dt^2-(1-2\Psi(\mathbf{x},t))\delta_{ij}dx^idx^j,
\end{equation}
where the induced gravitational potentials $\Psi(\mathbf{x},t)$ can be decomposed into the time-independent part $\Psi_c(\mathbf{x})$ and the oscillating part $\Psi_o(\mathbf{x})\cos(\omega t+2\alpha(\mathbf{x}))$. Given the mass $m$, the local energy density $\rho(\mathbf{x})$ and the local velocity $v(\mathbf{x})$ of DM particles, the two parts of $\Psi(\mathbf{x},t)$ can be obtained from Einstein equations~\cite{Khmelnitsky:2013lxt} 
\begin{align}
\nonumber
\Psi_o(\mathbf{x})=&\frac{\pi\hbar^2G\rho(\mathbf{x})}{m^2c^6}\\
=&6.48\times10^{-16}\left(\frac{\rho(\mathbf{x})}{0.4[{\rm GeV/cm^3}]}\right)\left(\frac{10^{-23}{\rm[eV]}}{mc^2}\right)^2,\\
\Psi_c(\mathbf{x})=&\frac{4\pi\hbar^2G\rho(\mathbf{x})}{m^2c^4v^2(\mathbf{x})}=4\times10^{6}\left(\frac{10^{-6}c^2}{v^2(\mathbf{x})}\right)\Psi_o(\mathbf{x}),\\
\omega=&\frac{2mc^2}{\hbar}=3\times10^{-8}{\rm[Hz]}\left(\frac{mc^2}{10^{-23}{\rm[eV]}}\right).
\end{align}
Therefore, a signal propagating in this metric will suffer a frequency shift
\begin{equation}
f_e-f_s=f_s(\Psi(\mathbf{x}_e,t_e)-\Psi(\mathbf{x}_s,t_s)),
\end{equation}
where the observables with the subscript $e$ are the ones detected at the Earth and the observables with the subscript $s$ are the ones detected at the source.
That is to say, this signal will suffer a frequency redshift $f_e<f_s$ when $\Psi(\mathbf{x}_e,t_e)<\Psi(\mathbf{x}_s,t_s)$.
At the position of the Earth, the velocity of DM is $v(\mathbf{x}_e)\sim10^{-3}c$ and the energy density of DM is $\rho(\mathbf{x}_e)=0.4{\rm[GeV/cm^3]}$~\cite{Nesti:2013uwa}. For a very nearby signal source ($d\sim{\rm100[pc]}$, $\Psi_o(\mathbf{x}_e)\approx\Psi_o(\mathbf{x}_s)$ and $\Psi_c(\mathbf{x}_e)\approx\Psi_c(\mathbf{x}_s)$), ultralight scalar DM with mass $m=10^{-23}[{\rm eV}/c^2]$ can induce a frequency shift $f_e-f_s\sim10^{-16}f_s$ in years. This tiny novel effect on the pulse frequency of the pulsar can be accumulated, and changes the pulse arrival time of the pulsar~\cite{Khmelnitsky:2013lxt}, then can be detected by the pulsar timing arrays~\cite{Kato:2019bqz}. 

If one want this simple frequency shift effect on the quasi-monochromatic GW signals from galactic WD binaries to be detected by LISA, the GW sources should be located in some DM clumps/subhalos~\cite{Bonaca:2018fek,Mirabal:2021ayb} where the energy density of DM is allowed to be $\rho(\mathbf{x}_s)\approx\frac{10^7M_{\odot}c^2}{10{\rm pc^3}}\approx10^8\rho(\mathbf{x}_e)$. 
In this paper, we will consider ultralight scalar DM with mass $m=n\times10^{-23}[{\rm eV}/c^2]$, velocity $v(\mathbf{x}_s)=1\times10^{-3}c$ and phase $\alpha(\mathbf{x}_s)=0$. Then we have $\Psi_o(\mathbf{x}_s)=10^8\Psi_o(\mathbf{x}_e)=\frac{6.48}{n^2}
\times10^{-8}$, $\Psi_c(\mathbf{x}_s)=10^8\Psi_c(\mathbf{x}_e)=\frac{2.59}{n^2}\times10^{-1}$ and $\omega=3n\times10^{-8}{\rm[Hz]}$. Then a GW signal with $f_s=f_0=1\times10^3{\rm[Hz]}$ from such DM clump suffer a frequency redshift
\begin{equation}
f_e-f_0=-f_0(\Psi_c(\mathbf{x}_s)+\Psi_o(\mathbf{x}_s)\cos(\omega t)).
\end{equation}
The first part $-f_0\Psi_c(\mathbf{x}_s)=-\frac{2.59}{n^2}\times10^{-4}{\rm[Hz]}$ is a time-independent frequency redshift. For $n>1$, it is not only negligible compared to $f_0=1\times10^3{\rm[Hz]}$ but also degenerate with $\mathcal{A}$, $\mathcal{M}$ and $d$ as shown in Eq.~(\ref{eq:am}). The second part $-f_0\Psi_o(\mathbf{x}_s)\cos(\omega t)=-\frac{6.48}{n^2}\times10^{-11}\cos(\omega t){\rm[Hz]}$ is a time-dependent frequency modulation with $\dot{f}_e=\frac{1.94}{n}\times10^{-18}\sin(\omega t){\rm[Hz^2]}$.
On the one hand, the detection of at least one oscillation of ultralight scalar DM during LISA's mission lifetime $T=4[{\rm year}]$ requires that $\omega$ should be larger than $5\times10^{-8}{\rm[Hz]}$ and $n$ should be larger than $1.67$; on the other hand, $\dot{f_e}>\sigma_{\delta\dot{f}}\dot{f}_0$ requires that $n$ should be smaller than $4.31$.
That is to say, LISA can detect ultralight scalar DM with mass $m=1.67\times10^{-23}-4.31\times10^{-23}[{\rm eV}/c^2]$ through the frequency modulation of quasi-monochromatic GW from galactic WD binaries located in DM clumps/subhalos. 

\begin{figure*}[]
\begin{center}
\subfloat{\includegraphics[width=1\textwidth]{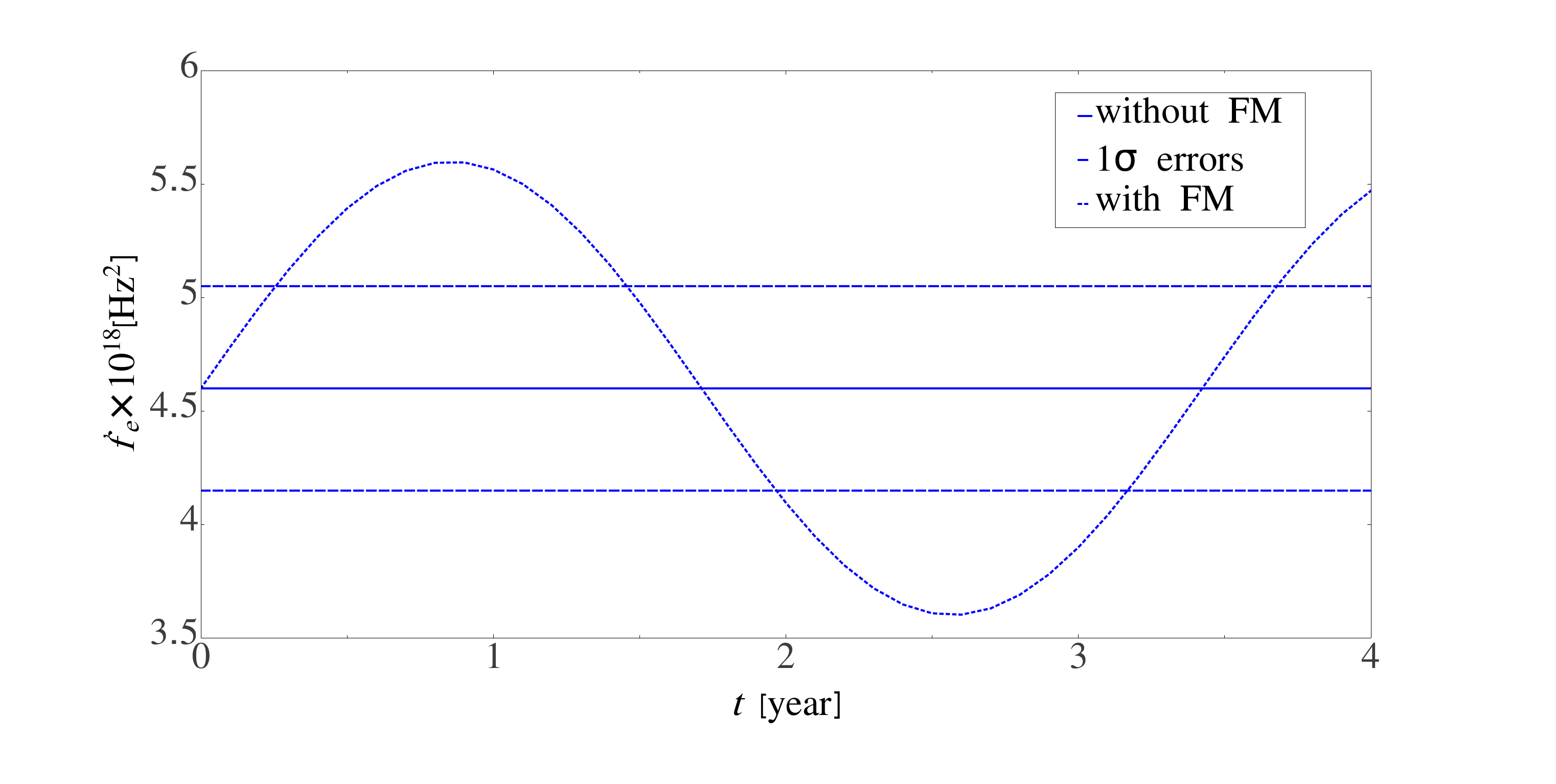}}
\end{center}
\captionsetup{justification=raggedright}
\caption{The evolution of $\dot{f_e}$ during LISA's mission lifetime. Without the frequency modulation by ultralight scalar DM, $\dot{f_e}$ just changes by $0.001\%$ during LISA's mission lifetime (blue solid line) and the measurement uncertainty of $\dot{f_e}$ are given by $\dot{f_e}(1\pm\sigma_{\delta\dot{f}})$ (blue dashed lines).
Taking the frequency modulation by ultralight scalar DM with mass $m=1.94\times10^{-23}[{\rm eV}/c^2]$ into consideration, the evolution of $\dot{f_e}$ is oscillating during LISA's mission lifetime (blue dotted curve).}
\label{fig:fdot}
\end{figure*}

In Fig.~\ref{fig:fdot}, the evolution of $\dot{f_e}$ only due to the GW radiation is the blue solid line, which just changes by $0.001\%$ during LISA's mission lifetime. The evolution of $\dot{f_e}$ due to the GW radiation and the frequency modulation by ultralight scalar DM with mass $m=1.94\times10^{-23}[{\rm eV}/c^2]$ is the blue dotted curve, which is oscillating across the measurement uncertainty of $\dot{f_e}$ (blue dashed lines).
This oscillating features are very distinguishable from the other simple chirping signals.
For example, as GW radiation drives the components of a galactic WD binary closer together, the effects of mass transfer and tidal forces will dominate the evolution of a negative $\dot{f_e}$ in $10^5$ years~\cite{Kremer:2017xrg}. The peculiar acceleration caused by a variation of the centre-of-mass velocity of a galactic WD binary will obtain a Doppler shifted $\dot{f_e}$~\cite{Xuan:2020xrr}.

\section{Summary and Discussion}
\label{sec:SD}

In this paper, inspired by the time-dependent frequency shift for the pulse signals of pulsars due to the oscillating pressure of the ultralight scalar DM, we propose a new novel detection method of the ultralight scalar DM. Similar to the pulse signals of pulsars, the quasi-monochromatic GW signals from galactic WD binaries are also can be considered as the probe to gather the oscillation information of the ultralight scalar DM during GW propagation. For $\rho(\mathbf{x}_s)\approx\rho(\mathbf{x}_e)=0.4{\rm[GeV/cm^3]}$~\cite{Nesti:2013uwa}, the time-dependent frequency shift for the pulse signals of pulsars can be accumulated in the arrival time of pulses, but the time-dependent frequency shift for the quasi-monochromatic GW signals from galactic WD binaries is very tiny. If we suppose that some WD binaries are located in the DM clumps/subhalos where $\rho(\mathbf{x}_s)\approx\frac{10^7M_{\odot}c^2}{10{\rm pc^3}}\approx10^8\rho(\mathbf{x}_e)$, the time-dependent frequency shift for the quasi-monochromatic GW signals from galactic WD binaries will be amplified accordingly. Compared to $\sigma_{\delta\dot{f}}$ estimated by the fisher information matrix, the frequency modulation of quasi-monochromatic GW from galactic WD binaries located in DM clumps/subhalos induced by the ultralight scalar DM with mass $m=1.67\times10^{-23}-4.31\times10^{-23}[{\rm eV}/c^2]$ will be detected by LISA.

There are two caveats. The first one is that we have supposed that some galactic WD binaries with chirp mass measured by LISA are located in the DM clumps/subhalos. Given the number of such WD binaries (about 1000) and the number of the DM clumps/subhalos (about 100) in the Milky Way, this assumption seem to be reasonable. But in reality we don't know the true distribution of WD binaries in the Milky Way. We also don't know whether or not the WD binaries are excluded from the DM clumps/subhalos. The second one is the conflict between the concept of ULDM and the concept of DM clump/subhalo. The former one is introduced to suppress the sub-galactic structures, but the latter one is just the sub-galactic structure. We don't know how ULDM forms the DM clumps/subhalos. Here we just assume that there is an additional local potential well at the location of DM clumps/subhalos.
All in all, the detection of the GW frequency modulation can also help us to investigate the DM clumps/subhalos in the Milky Way.

\begin{acknowledgments}
We acknowledge the use of HPC Cluster of Tianhe II in National Supercomputing Center in Guangzhou. Ke Wang is supported by grants from the National Key Research and Development Program of China (grant No. 2021YFC2203003), grants from NSFC (grant No. 12005084 and grant No.12247101) and grants from the China Manned Space Project with NO. CMS-CSST-2021-B01.
\end{acknowledgments}

\end{document}